\documentclass[groupedaddress,notitlepage,article]{revtex4-1}

\usepackage{graphicx,psfrag}
\usepackage{epsfig}
\usepackage{amsmath}
\DeclareGraphicsExtensions{.eps}

\begin{document}

\title{Phonon engineering with superlattices: Generalized nanomechanical potentials}

\author{O. Ort\'iz, M. Esmann, and N. D. Lanzillotti-Kimura}
\email[]{daniel.kimura@c2n.upsaclay.fr}
\affiliation{Centre de Nanosciences et de Nanotechnologies (C2N), CNRS, Universit\'e Paris-Sud, Universit\'e Paris-Saclay, 10 Boulevard Thomas Gobert, 91120 Palaiseau, France}

\begin{abstract}
Earlier implementations to simulate coherent wave propagation in one-dimensional potentials using acoustic phonons with gigahertz-terahertz frequencies were based on coupled nanoacoustic resonators. Here, we generalize the concept of adiabatic tuning of periodic superlattices for the implementation of effective one-dimensional potentials giving access to cases that cannot be realized by previously reported phonon engineering approaches, in particular the acoustic simulation of electrons and holes in a quantum well or a double well potential. In addition, the resulting structures are much more compact and hence experimentally feasible. We demonstrate that potential landscapes can be tailored with great versatility in these multilayered devices, apply this general method to the cases of parabolic, Morse and double-well potentials and study the resulting stationary phonon modes. The phonon cavities and potentials presented in this work could be probed by all-optical techniques like pump-probe coherent phonon generation and Brillouin scattering.
\end{abstract}

\maketitle

\section{Introduction}
Nanophononics addresses the control of acoustic phonons in solid state structures with engineered acoustic impedance modulations\;\cite{volz_nanophononics_2016,balandin_nanophononics_2005,balandin_phonon_2007,lanzillotti-kimura_nanophononic_2010,lamberti_nanomechanical_2017}. Commonly studied nanoacoustic devices include phonon mirrors, filters and resonant cavities to shape the interaction of phonons at the GHz-THz frequency scale with both light and electronic states. Applications include fast modulators in semiconductor lasers\;\cite{bruggemann_laser_2012}, novel approaches for the generation of THz radiation\;\cite{armstrong_observation_2009} and the nanomechanical characterization of biological tissue\;\cite{Dehoux_optoacoustic_2012,Gusev_advances_2018}. Optical tools such as ultrafast pump-probe spectroscopy and inelastic Brillouin scattering have enabled the study of phononic spectra, temporal dynamics and coherence properties on the nanoscale\;\cite{Imade_gigahertz_2017,Berte_acoustic_2018,Xu_all_2018,Stoll_time_2015,rozas_lifetime_2009,thomsen_coherent_1984,thomsen_surface_1986,huynh_subterahertz_2006}.  This paved the way to establish nanoacoustics also as a platform for the simulation of wave dynamics\;\cite{bruchhausen_acoustic_2018}. In contrast to optical platforms, nanophononics features the particular advantage of a slow speed of propagation compared to light and a long coherence length in the range of hundreds of micrometers\;\cite{huynh_subterahertz_2006} at wavelengths in the 10 nm range. Therefore coherent propagation of phonons can be studied in quasi-infinite systems in which wave dynamics can be optically probed on timescales well below the mechanical oscillation period.                    

To mimic the dynamics of electrons in potentials using acoustic phonons, a band structure is necessary. A well-established building block in the engineering of acoustic phonons is the acoustic nanoresonator based on the acoustic counterpart of an optical Fabry-Perot cavity\;\cite{trigo_confinement_2002-1,huynh_subterahertz_2006}.  Most of the earlier approaches to implement potentials are based on engineering bands arising from coupled nanoacoustic cavities\;\cite{kimura_phonon_2007}, i.e. the phononic equivalent of the coupled resonator optical waveguides (CROWs)\;\cite{yariv_guided_1999}. Such devices have been used to e.g. mimic wave dynamics in Wannier-Stark ladders showing Bloch oscillations\;\cite{lanzillotti-kimura_bloch_2010} or topological effects in polyacetylene\;\cite{esmann_topological_2018-4}. In contrast to coupled cavities, another approach to study wave dynamics in effective acoustic potentials is to shape a local band structure along a single periodic multilayer. Very recently, an acoustic cavity was reported based on the adiabatic periodicity breaking of a superlattice\;\cite{lamberti_nanomechanical_2017}, in analogy to a potential well. Exploiting the symmetry properties in periodic superlattices, topological interface modes have also been recently reported\;\cite{esmann_topological_2018-1,esmann_a_2018}.

Here, we generalize the adiabatic tuning of the period thickness in a superlattice for the implementation of effective one-dimensional potentials. Up to now, the realization of effective phononic potentials was predominantly based on coupled resonant cavities exploring the thight-binding physics of phonons tunneling between sites on a lattice\;\cite{kimura_phonon_2007,yariv_guided_1999,lanzillotti-kimura_bloch_2010,esmann_topological_2018-4}. In contrast, here we explore the physics of nearly-free electron models based on a single superlattice. We demonstrate that potential landscapes can be tailored with great versatility using significantly thinner structures than previously reported, rendering even the implementation of complicated effective potentials experimentally feasible. We apply this general method to the cases of parabolic, Morse and double-well potentials and study the resulting stationary phonon modes.

The paper is organized as follows: Section 2 presents a general theoretical framework to engineer acoustic cavities based on adiabatic changes in the local acoustic band structures. We present a first case mimicking the confinement of an electron and a hole in a quantum well. In Section 3, the use of these tools is extended for the implementation of effective parabolic, double-well and Morse potentials and their stationary acoustic modes are calculated. Section 4 presents the conclusions and perspectives of this work.

\section{Acoustic potential wells}

Potential wells for acoustic phonons can be implemented based on periodic multilayered structures. The theoretical framework to engineer these effective potentials departs from the dispersion relation of phonons in a periodic superlattice\;\cite{cardona_light_1983} with each unit cell composed of two layers. Considering $AlAs$ and $GaAs$ as the materials for these layers, we parametrize this dispersion relation as
\begin{eqnarray}
	\cos(k d)=&\cos&\left(\frac{\omega D}{f_d}\right)\nonumber\\&-&\frac{\epsilon^2}{2}\sin\left(\frac{\omega r D}{f_d}\right)\sin\left(\frac{\omega (1-r)D}{f_d}\right)
\end{eqnarray}
with
\begin{equation}
	\epsilon=\frac{Z_{GaAs}-Z_{AlAs}}{(Z_{GaAs}Z_{AlAs})^{1/2}}
\end{equation}
Here, $k$ is the phonon quasi-momentum, $d=d_{AlAs}+d_{GaAs}$ the geometric unit cell thickness (i.e. structure period) composed of two layers and $\omega$ the angular frequency. $Z_{GaAs} = \rho_{GaAs} v_{GaAs}$ and $Z_{AlAs} = \rho_{AlAs} v_{AlAs}$ are the acoustic impedances of the two materials where $\rho_{AlAs}$ and $v_{AlAs}$ ($\rho_{GaAs}$ and $v_{GaAs}$) are the mass density and the speed of sound, respectively. We denote the acoustic path length of a unit cell $D=D_{AlAs}+D_{GaAs}$ with the contributions $D_{AlAs}=d_{AlAs}f_d/v_{AlAs}$ and $D_{GaAs}=d_{GaAs}f_d/v_{GaAs}$ from the two layers, that are measured in units of wavelength at a design frequency $f_d$. We furthermore introduced the parameter 
\begin{equation}
	r=\frac{D_{AlAs}}{D}
\end{equation}
which describes the relative contribution of the $AlAs$ layer to the overall acoustic path length of the unit cell. 
\begin{figure}[ht!]
	\centering 
	\includegraphics[width=8.6cm]{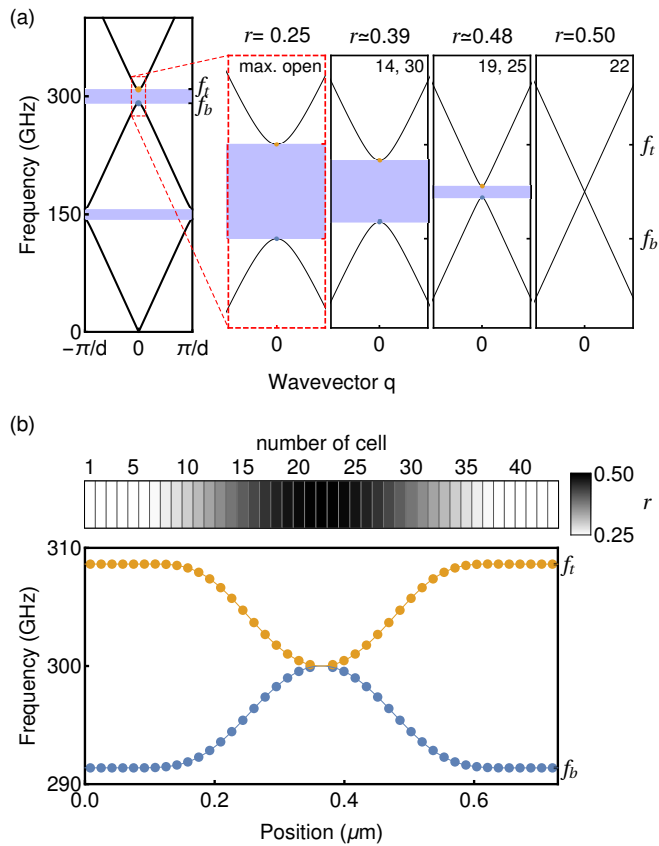}
	\caption{(a) Left: acoustic band diagram of a superlattice with its second minigap maximally opened. The first two acoustic minigaps at 150 GHz and 300 GHz are highlighted in light blue. They correspond to the first minigaps at Brillouin zone edge and center, respectively. $f_b$ and $f_t$ mark the bottom and top edge of the second minigap. Right: zoom-in the acoustic band structure around the second acoustic minigap (red dashed rectangle) associated to the unperturbed cells, cells 14 and 30, 19 and 25, and 22 respectively. $r$, the ratio of acoustic thickness of $AlAs$, is depicted on top of each frame. (b) Top: Schematic of an adiabatic perturbation induced in a periodic structure of 43 unit cells. The grayscale indicates the value of $r$ for each cell. Its modulation follows a $\cos^2$ shape between cells 17 and 49 with an amplitude of 0.25 while remaining constant at 0.25 in the rest of the structure. Bottom: Local acoustic band structure representing the minigap edges for each cell as a function of position. Each vertically aligned pair of dots correspond to the minigap edges associated to each cell and plotted at each cell center. The curves mark the continuous trajectory targeted for each edge.}
	\label{adiabColumnFig}
\end{figure}
This dispersion relation is plotted in the leftmost frame of Fig. \ref{adiabColumnFig}(a) for a $GaAs$/$AlAs$ superlattice with $f_d=300\,\rm{GHz}$, $D=1$ and $r=0.25$ i.e. a phonon at the design frequency $f_d$ acquires a propagation phase of $2\pi$ upon traversing one unit cell and a phase of $\pi/2$ upon traversing one $AlAs$ layer. We observe three acoustic bands separated by two acoustic minigaps around 150 GHz and 300 GHz in which only evanescent phonons are solutions to the wave equation. Since $D$ is chosen as an integer multiple of $1/2$, $f_d$ lies at the center of a minigap, in this case the second one. The red dashed rectangle in the second minigap marks the region of the band structure for which zoom-ins are presented in the right part of panel (a) for different values of the parameter $r$. Importantly, we observe that the spectral width of the minigap critically depends on the value of $r$, i.e. on the relative acoustic thickness of the two layers constituting the unit cell\;\cite{cardona_light_1983,esmann_topological_2018-1}.  Starting from a maximally opened band gap at $r=0.25$ (first frame), the gap completely vanishes for a value of  $r=0.5$, i.e. when the two material layers have equal acoustic thickness (this holds for the second minigap, for higher order, several such points can be identified). We use this dependence to construct a first example of an effective phonon potential by slowly (adiabatically) varying the parameter $r$ inside a multilayer structure. Departing from a 43 unit cells periodic structure with $f_d=300\,\rm{GHz}$, $D=1$ and $r=0.25$, the potential is designed by modulating $r$ following a $\cos^2$ pattern\;\cite{lamberti_nanomechanical_2017} along the 33 central unit cells (see top part of Fig. \ref{adiabColumnFig}(b)). There, $r$ first increases up to $r=0.5$ at the central cell and then changes back to the initial value of $r=0.25$. By doing so, the width of the second minigap undergoes a closing and reopening along the structure. The term adiabatic refers to the condition that the considered minigap shows a large overlap between consecutive unit cells. In this work, our structures have typical overlaps beyond 93\%. The corresponding evolution of the upper and lower band edges is shown in the bottom of Fig. \ref{adiabColumnFig}(b). Here, we adopt the notion of a 'local band structure'\;\cite{lamberti_nanomechanical_2017,kimura_phonon_2007,Lermer_Bloch_2012,malpuech_picosecond_2001,agarwal_photon_2004}, i.e. each unit cell of the structure is assigned the bands of an infinite superlattice having the same structural parameters. 

\begin{figure}[t]
\centering 
\includegraphics[width=8.6cm]{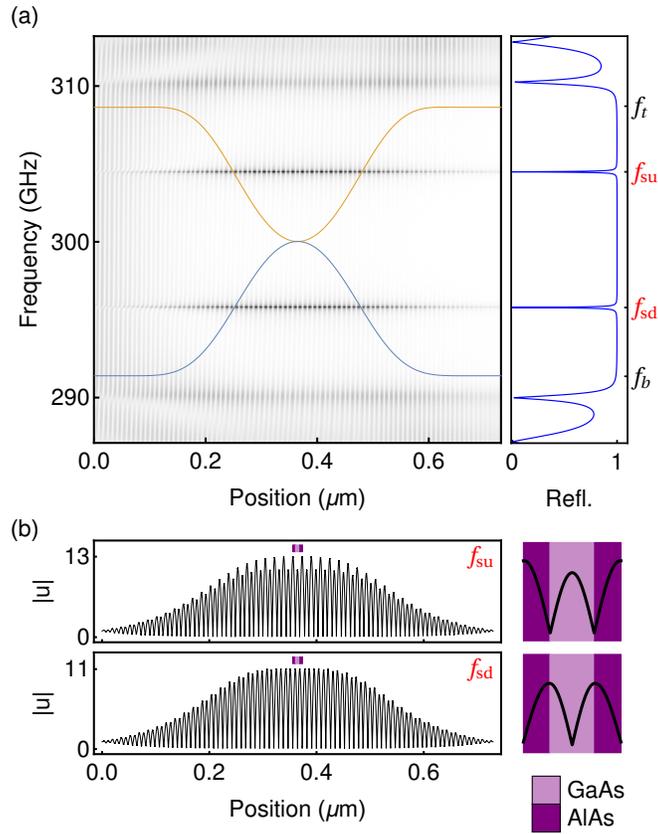}
\caption{(a) Right: acoustic reflectivity of an adiabatic resonator created by introducing a modulation in the ratio of $AlAs$ per cell in a periodic structure. $f_b$ and $f_t$ mark the maximally open stopband limits while $f_{sd}$ and $f_{su}$ the frequencies of the confined modes. Left: displacement distribution as a function of position and frequency for phonons propagating from left to right. (b) Left: normalized displacement profile of the confined modes. Right: for each mode, a zoom-in the profile at the central cell is show.}
\label{newSymFig}
\end{figure}

To illustrate how this structure acts as an effective phononic potential, we calculate its acoustic reflectivity spectrum shown in the right frame of Fig. \ref{newSymFig}(a) using a transfer matrix formalism. The displayed spectrum presents a central high reflectivity band containing two sharp reflectivity dips marked $f_{sd}$ and $f_{su}$. Outside this region the reflectivity shows Bragg oscillations, a common feature in superlattices\;\cite{Ezzahri_coherent_2007,trigo_confinement_2002-1}. The high reflectivity region corresponds to the frequency range of the maximally open minigap $f_b<f<f_t$, where the structure acts as a distributed Bragg reflector (DBR). The two dips correspond to two resonances of the structure. Their mechanical displacement profiles $u(z)$ along the superlattice are plotted in Fig. \ref{newSymFig}(b). Both states are confined to the modulated minigap region with exponentially decaying tails inside the unmodulated outer parts of the structure. In an earlier report, an adiabatic nanoacoustic cavity was presented\;\cite{lamberti_nanomechanical_2017} in which the width of a bandgap was kept constant and an effective potential well was obtained by locally shifting the gap along the structure. In contrast to this case, we here kept the central energy of the gap constant and changed its width along the structure. We furthermore compute the mechanical displacement within the structure upon an incident plane wave of unit amplitude as a function of frequency and position. The plot is superimposed with the $\cos^2$ evolution of the local band edges (blue and orange lines) in Fig. \ref{newSymFig}(a) (left frame). The modulation of $r_{AlAs}$ in the central region closes and re-opens the local minigap producing a wasp-waist shaped local band structure. The two confined modes at $f_{sd}$ and $f_{su}$ are evidenced by a strongly enhanced mechanical response in the modulated central part of the superlattice. The superlattice very much resembles the behavior of a quantum-mechanical potential well with bound, localized solutions at energies below the band edge and a continuum of propagating solutions for energies outside the maximally open band gap. In analogy to a quantum mechanical potential well, the number of bound states increases if the well is widened or deepened. Interestingly, the potential well studied here supports two confined states, one in its upper convex dip at frequency $f_{su}$, the other in the lower concave part at a frequency $f_{sd}$. While both states present similar envelopes, their overall spatial symmetry of the carrier is opposite (see the zoom-in of $|u|$ along the central cell at the right of each frame in panel (b)). This is a direct consequence of the different spatial symmetries of the Bloch modes at the upper and lower band edge\;\cite{zak_symmetry_1985,xiao_surface_2014,esmann_topological_2018-1} (marked yellow and blue in panel (a), respectively). These states are the equivalent of a bound electronic state and a bound hole. This observation represents a new aspect of adiabatic cavities compared to earlier works where only single-sided potential wells were considered\;\cite{Lermer_Bloch_2012,lamberti_nanomechanical_2017}. In the rest of the paper we will extend the concept of phononic band engineering to generalized acoustic potentials.

\section{Effective Potentials}
We have shown how to achieve a double-sided acoustic potential well by closing and reopening a minigap. Extend this approach for general potentials requires to modulate both the spectral width of a phononic minigap and its central frequency. The latter is achieved by scaling the overall thickness of the unit cell. To this end, we re-write the dispersion relation of a bilayer superlattice as
\begin{eqnarray}
	1=&\cos&\left(\frac{2 \pi f_b}{f_d}\right)\nonumber\\
	&-&\frac{\epsilon^2}{2}\sin\left(\frac{2 \pi f_b r}{f_d}\right)\sin\left(\frac{2 \pi f_b (1-r)}{f_d}\right) \label{dispRel3}
\end{eqnarray}
Here we have set $k=0$ because we target the second minigap at the center of the Brillouin zone. We have furthermore set $f=f_b$ and $D=1$. This allows us to shape the local band structure by placing its bottom edge at a fixed value $f_b$ while controlling the top edge at will through varying $f_d$ and solving Eq. \ref{dispRel3} for $r$. Next, we will illustrate this approach by designing three structures mimicking a parabolic, a Morse and a double-well potential.

	\subsection{Parabolic potential}

	As a first case we study a parabolic potential. This potential is usually associated to a harmonic oscillator with equidistant energy levels. While the quadratic dispersion relation of electrons indeed results in eigenvalues following a linear dependence with respect to the mode number $n$, the linear dispersion relation of phonons in contrast results in a dependence proportional to $n^{2/3}$ mimicking the case of relativistic electrons or light. In the optical domain, similar effective potentials have been demonstrated using two-dimensional photonic crystal structures\;\cite{crosnier_integrated_2016,combrie_integrated_2017}. In nanoacoustics a similar potential has been proposed using coupled acoustic nanocavities\;\cite{kimura_phonon_2007,bruchhausen_acoustic_2018}.

	Using Eq. \ref{dispRel3}, we design an acoustic structure mimicking a parabolic potential of the form
	\begin{equation}
		V(x)=V_0\;x^2. 
	\end{equation}
	The variable $x$ is a unitless, normalized position varying from $-1$ to $1$  from the leftmost to the rightmost unit cell of the structure. The amplitude of the potential is chosen as $V_0=f_t-f_b$ such that its value ranges from a maximally opened minigap at the first and last unit cell to a closed gap at the center. Note that both $V(x)$ and $x$ vary in discrete steps from cell to cell along the structure. To obtain an effectively smooth potential, we choose a structure composed of $N=101$ cells and numerically solve the $N$ equations to obtain the parameter $r$ defining the local band structure for each cell. 
		\begin{figure}[t]
			\centering 
			\includegraphics[width=8.6cm]{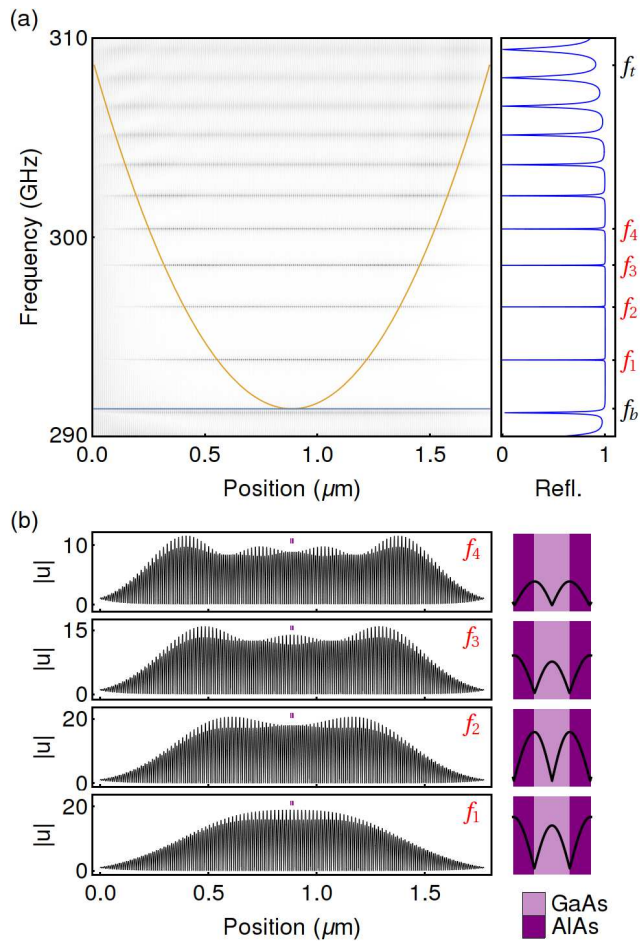}
			\caption{(a) Right: acoustic reflectivity for the parabolic potential structure. $f_b$ and $f_t$ mark the DBR stopband limits while $f_{1}$--$f_{4}$ the first four confined modes. Left: displacement distribution as a function of position and frequency for phonons propagating from left to right. On top the local band structure is plotted. (b) Left: normalized displacement profile of the first four confined modes. Right: for each mode, a zoom-in the profile at the central cell is show.}
			\label{newParabFig}
		\end{figure}		
	
	Figure \ref{newParabFig}(a) (right frame) presents the acoustic reflectivity for this structure, where we observe a series of dips associated to the confined modes in the acoustic potential. The modes labeled $f_b$ and $f_t$ indicate the superlattice stopband edges for the maximally opened minigap. The modes labeled $f_1$ to $f_4$ correspond to the first four bound states. The left frame in panel (a) shows the local band structure on top of the displacement distribution color map calculated in the same manner as in Fig.\ref{newSymFig} (a). Modes $f_1$ to $f_4$ are localized inside the parabola similarly to what was observed in the previously discussed potential well. The unequal spacing between modes observed follows the expected $n^{2/3}$ dependence. Figure \ref{newParabFig}(b) shows the displacement profile corresponding to an incident plane wave of unit strength for each of the first four modes. They are mainly localized in the center of the structure with two evanescently decaying tails on the sides. Their envelopes exhibit $n$ maxima for the $n$-th confined mode. For the first four modes, notice that the length of the evanescent tails decreases with increasing order. This can be explained by considering that the penetration depth for phonons at frequencies within a DBR stopband not only depends on the minigap bandwidth but also on the spectral position within the gap. For a given width of the gap, the evanescent decay length of a mode is the shortest if it lies at the bandgap center and diverges when approaching the gap edges. Therefore, the decaylength of the modes in the parabolic potential decreases with ascending order since the modes appear closer to the bandgap center. For further analysis, each confined mode profile in Fig.\ref{newParabFig}(b) includes a zoom-in of its central cell depicted at the right of each frame. The central cell position and size is also represented as an inset within each frame. Similarly to the acoustic potential well presented in the previous section, the fundamental mode confined by the top edge of the second acoustic minigap shares this edge's mode symmetry (symmetric, represented by the yellow curve marking the top edge of the local band structure). For the higher order modes we however observe alternating symmetry properties in agreement with the well-known eigenstates of the quantum-mechanical harmonic oscillator\;\cite{Kittel2004}.

	\subsection{Morse potential}
	
		The proposed method enables us to implement potentials that are asymmetric in space such as the Morse potential. This potential is used to model interatomic interactions in diatomic molecules. It presents a well shape but it distinguishes itself from the quantum harmonic oscillator by an asymptotic limit in the potential energy on one side. By doing so, it includes the effect of chemical bond breaking that the quantum harmonic oscillator does not consider. The resulting modes of such a potential are either bound below the asymptotic energy limit or unbound above it.
		
		The Morse potential as a function of a radial coordinate $r$ can be expressed as
		\begin{equation}
			V(r)=V_0 \left\lbrace\frac{1 - \exp[a\;(r_e-r)]}{1 - \exp[a\;r_e]}\right\rbrace^2
		\end{equation}
		with the parameters $V_0$, $a$ and $r_e$ determining the depth, width and equilibrium position of the potential, respectively. The variable $r$ is a unitless, normalized position along the structure varying from $0$ to $1$ from the leftmost to the rightmost unit cell. We set $a\approx3.95$ and $r_e=0.24$ while choosing $V_0=f_t-f_b$, i.e. the potenial spans the full range from a completely closed to a fully opened acoustic minigap. 
		
\begin{figure}[t]
			\centering 
			\includegraphics[width=8.6cm]{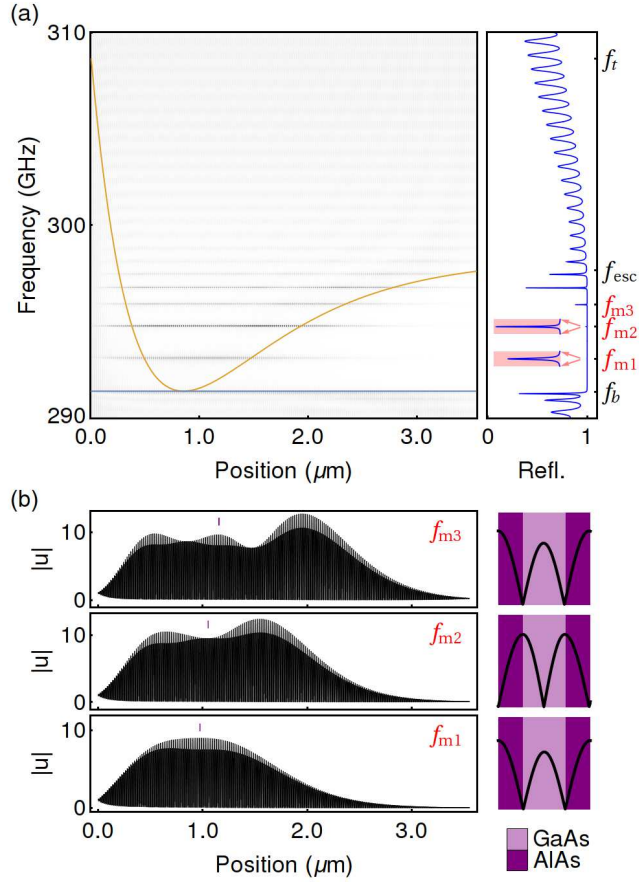}
			\caption{(a) Right: acoustic reflectivity for the Morse potential structure. $f_b$ and $f_t$ mark the DBR stopband limits while $f_{m1}$, $f_{m2}$ and $f_{m3}$ the first three confined modes. Inset: zoom-in the dips around $f_{m1}$ and $f_{m2}$. Left: displacement distribution as a function of position and frequency for phonons propagating left to right. Local band structure plotted on top. (b) Left: normalized displacement profile of the first three confined modes. Right: for each mode, a zoom-in the profile at the central cell of the mode is show.}
			\label{newMorseFig}
		\end{figure}

	Figure \ref{newMorseFig}(a) (right frame) presents the acoustic reflectivity of such a structure composed of 201 unit cells. As before, $f_b$ and $f_t$ correspond to the edges of the maximally opened minigap. Two main bands can be distinguished; a high reflectivity band where three modes appear (labelled $f_{m1}$, $f_{m2}$ and $f_{m3}$) and a higher energy band with reflectivity oscillations. In contrast to the parabolic potential, which showed  marked reflectivity dips, the modes of the Morse potential are associated to shallow reflectivity dips. This can be explained by considering that the structure is asymmetric, similar to an unbalanced Fabry-Perot resonator\;\cite{Winter_selective_2007}, i.e. a structure effectively composed of two mirrors with unequal reflectivity. The imbalance causes predominant decay of the confined modes to one side and hence limits the achievable minimum reflectivity upon interference of multiple internal reflections. The fact that the imbalance is reduced while going to higher frequencies, together with the change in penetration depth for different frequencies (as explained for the parabolic structure) causes the confinement to increase with increasing order of the mode. The left frame in Fig. \ref{newMorseFig}(a) shows the local band structure on top of the displacement color map in response to a plane wave of unit strength incident from the left. We see the confined modes corresponding to the bounded states within the dip of the Morse potential for frequencies below $f_{esc}$.  Above this energy, a series of unbounded modes appear corresponding to the oscillations in the reflectivity plotted in Fig. \ref{newMorseFig}(a) (right frame). 
	Finally, Fig. \ref{newMorseFig}(b) shows the displacement profile of the first three confined modes. Comparing these with those of Fig. \ref{newParabFig}(b), again the number of maxima in the envelope corresponds to the order of the mode. In the case shown here, the left and right evanescent tails are different. Once again, we see the symmetric/anti-symmetric alternation between consecutive confined modes as depicted in the central cells at the right of each frame in Fig. \ref{newMorseFig}(b). Since the structure is not symmetric, we define the central cell for each confined mode as the cell with the central extreme of the mode envelope.

	\subsection{Double-well potential}
	
		Based on the same design principles, we engineer the phononic equivalent of a hydrogen molecule. In the context of quantum mechanics, it is the result of the hybridization of degenerate eigenstates of two individual quantum wells, similar to the one presented in section A. This double well potential thus has eigenstates that extend over two well-defined regions of space separated by a tunneling barrier. By modifying the barrier, it is possible to control the coupling strength and hence the energy splitting of the hybridized modes. Note that this type of potential can also be reasoned with a tight binding approach, i.e. each of the two atoms is represented by an individual parabolic potential, and the hopping term by an effective transmission through the barrier \;\cite{kimura_phonon_2007}. Previous realizations of acoustic molecules have been based on the coupling of pairs of cavities\;\cite{bruchhausen_acoustic_2018}. The usage of a double-well structure presents its own advantages. By customizing the shape of the double well control on the envelope of the confined modes is gained. Additionally, each isolated phononic atom supports multiple modes besides the fundamental one. In this way, the energy and the splitting of the hybridized modes can be engineered. The optical counterpart of this systems, known as photonic molecules\;\cite{bayer_optical_1998}, appear as promising platform for the development of quantum technologies and could inspire future phononic technologies\;\cite{dousse_ultrabright_2010,zhang_electronically_2019}.
		
		The potential energy as a function of the parameter $z$ for a double-well potential can be expressed as
		\begin{equation}
			V(z)=V_0(\lambda \; z^4 - k\;z^2)+k^2/4\lambda
		\end{equation}
		where $\lambda$ and $k$ are both positive parameters determining the position and depth of both wells. The variable $z$ is the unitless, normalized position along the structure varying from $-1$ to $1$ from the leftmost to the rightmost unit cell. We choose $\lambda\approx3.58$ and $k=3.38$ such that it spans values from $0$ to $V_0=f_t-f_b$.

			\begin{figure}[t]
				\centering 
				\includegraphics[width=8.6cm]{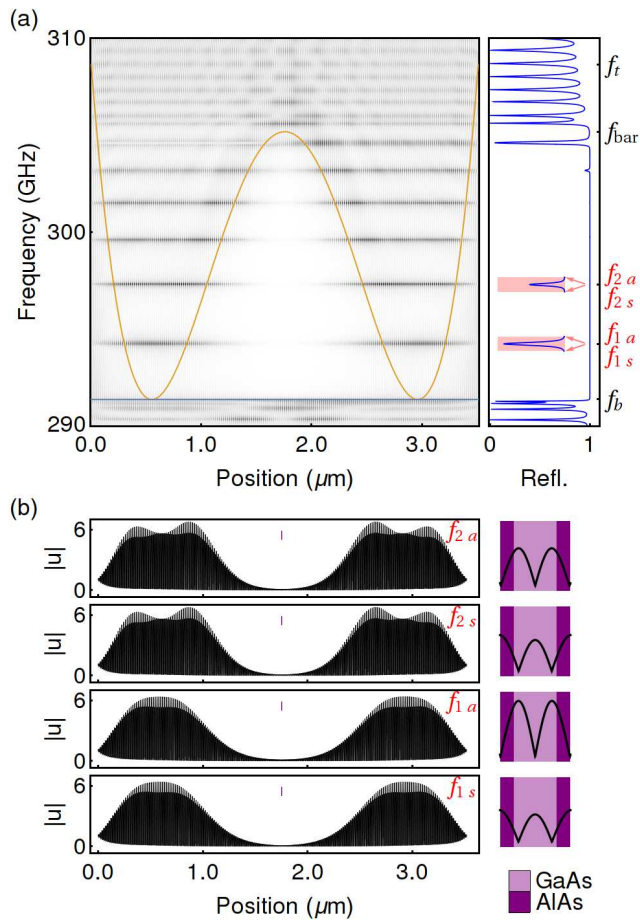}
				\caption{(a) Right: acoustic reflectivity for the double-well potential structure. $f_b$ and $f_t$ mark the DBR stopband limits while $f_{1s}$, $f_{1a}$, $f_{2s}$ and $f_{2a}$ the first four confined modes. Inset: zoom-in the shallow dips around resonant frequencies. Left: displacement distribution as a function of position and frequency for phonons incident to the structure from both ends with a $\pi/2$ dephase between each other. Local band structure plotted on top. (b) Left: normalized displacement profile of the first four confined modes. Right: for each mode, a zoom-in the displacement profile at the central cell.}
				\label{newDwFig}
			\end{figure}

		Figure \ref{newDwFig}(a) (right frame) shows the acoustic reflectivity of a double-well structure built of $201$ unit cells. The top edge frequency at the center hump is labeled as $f_{bar}$, while $f_b$ and $f_t$ correspond to the edges of the maximally opened minigap. The oscillations above $f_{bar}$ correspond to modes localized in the full structure. Below $f_{bar}$ a high reflectivity region with shallow dips appears. Insets show a zoom-in the regions around the first two dips. By performing a numerical search for the  resonant frequencies in the neighborhood of both dips, four instead of only two solutions are found. These are labeled $f_{1s}$, $f_{1a}$, $f_{2s}$ and $f_{2a}$. Figure \ref{newDwFig}(a) (left frame) shows the local band structure on top of the displacement color map. For this case, in order to visualize modes at both wells with similar amplitudes, two plane waves are considered entering the structure coming from both sides with a $\pi/2$ phase between them. Because of this, interference between the counterpropagating modes produces a color map that does not fully preserve the symmetry of the potential. 
		We can see how bound modes below $f_{bar}$ are simultaneously localized in both potential wells with very little amplitude in the central hump of the structure. This central hump acts as a potential barrier between each well. Tunneling through it, acoustic phonons propagate from one well to the other. Consequently, increasing or reducing the barrier will increase or reduce the coupling between the wells. Figure \ref{newDwFig}(b) shows the displacement profiles of the first four resonant modes. Each pair of them has a very similar shape of the envelope. A more detailed analysis however evidences the hybridization into symmetric and anti-symmetric solutions. At the right side of each frame a zoom-in the displacement profile at the central cell is presented.  As before, we can see how the first mode shares the symmetry of its enclosing edges, while the rest alternate exchanging nodes with antinodes. As in the other potentials, the number of maxima in the mode envelopes scales with the order of the mode pair. In resemblance with the Morse case, we can see the envelope not being symmetric with respect to each well but increasing towards the center of the structure. This, together with the shallow reflectivity dips, results from each confinement region (well) being surrounded by unbalanced reflective regions as in the Morse potential case.

\section{Discussion and Conclusions}

We have demonstrated how adiabatic deformations of nanoacoustic superlattices can be used for the implementation of arbitrary effective potential landscapes for longitudinal acoustic phonons. First, by introducing the concept of a local band structure, we have designed an adiabatic potential well resonator, and shown how it confines acoustic modes mimicking electron and hole states. Secondly, we have parametrized the dispersion relation of bilayer superlattices isolating parameters determining the minigaps bandwidths and positions. Using this parametrization, three potential landscapes have been studied: a parabolic, a Morse and a double-well potential. While the main features of the electronic counterparts are recovered, effective phonon potentials present significant novel aspects to remark.
The spatial profile of the confined modes (in particular its decay length towards the exterior) depends on the position and frequency dependent penetration depth as well as the boundary conditions. For the Morse and double-well potential the dips in the reflectivity curve corresponding to the confined modes are shallow as a consequence of confinement between unbalanced effective phonon mirrors. Finally, the double-well potential highlights the role of the symmetries of the spatial mode profiles in hybridized mode solutions. In all three cases, the spectrum of confined modes is fundamentally different from the standard electronic counterpart, since the phonon dispersion relation in a bulk material is linear rather than parabolic. The systems described here correspond to the dynamics of quasirelativistic particles in potential wells which have a linear dispersion\;\cite{kimura_phonon_2007,Hall_energy_2001}.

The potentials are implemented based on adiabatic deformations of the unit cell structure along a periodic superlattice of typically one hundred cells. From a practical point of view, such lattices can be experimentally realized in molecular beam epitaxy (MBE) growth\;\cite{Kimura_nanowave_2006}. Coherent phonon generation and detection experiments could be performed to experimentally probe the time dynamics in these structures, accessing not only the spatial but also the temporal evolution of the wave function. In comparison, relying on coupled cavities to study phonon dynamics rather than single superlattices, an equivalent implementation of the potentials presented in this work would require materials with higher impedance contrast or very thick samples, which are much more demanding to grow\;\cite{bruchhausen_acoustic_2018}. 

A necessary condition in both approaches is that the system allows a description in terms of local band structures. Here, we ensured this by keeping the overlap between the second minigap of consecutive unit cells sufficiently large. As a consequence, the shape of the potential landscape to be mimicked will impose a minimum number of layers needed depending on its maximum gradient. By implementing potentials with a single superlattice it is thus possible to experimentally realize potentials that are unrealistic with coupled nanocavities. Moreover, key results that were already achieved based on the tight-binding approach with coupled cavities\;\cite{kimura_phonon_2007}, such as phonon molecules and Bloch oscillations in Wannier-Stark ladders, could be also implemented through the nearly-free electron approach developed here\;\cite{bruchhausen_acoustic_2018}.

The materials $GaAs$ and $AlAs$ used in our simulations present the unique feature that their acoustic impedance contrast and index of refraction contrast are nearly the same. As a result, cavities simultaneously confining both sound and light have been achieved. Exploiting this feature, the design procedure for general potential landscapes introduced here could be extended to optomechanical systems with near-perfect colocalization of sound and light\;\cite{Fainstein_strong_2013,Arregui_anderson_2018}.

Another fundamental property of acoustic superlattices that has not been exploited here are the Bloch mode symmetries at the minigap edges. It has been shown\;\cite{xiao_geometric_2015} that an inversion of these symmetries can be achieved by closing and re-opening a minigap, inducing a topological phase transition and hence allowing the construction of topological interface states\;\cite{esmann_a_2018,esmann_topological_2018-1,esmann_topological_2018-4,Maire_optical_2018}. This approach could for example be combined with the adiabatic resonator presented in Fig.\ref{newSymFig}, resulting in an additional confined state between the two modes of the well, i.e. mimicking a zero-energy state between electron and hole.
Overall, we presented a series of acoustic devices allowing the control of the propagation of longitudinal acoustic phonons that are achievable through standard MBE growth and can be tested using optical pump-probe and Brillouin scattering schemes.

\begin{acknowledgments}
The authors thank F.R. Lamberti, P. Senellart and L. Lanco for fruitful discussions. The authors acknowledge funding by the European Research Council Starting Grant No. 715939, Nanophennec; ME acknowledges funding by the Deutsche Forschungsgemeinschaft (DFG, German Research Foundation) – Project 401390650.
\end{acknowledgments}

\end{document}